\documentclass[a4paper,11pt]{article}
\usepackage{jheppub} 
\usepackage{lineno}
\usepackage{subcaption}
\usepackage{mathtools}
\usepackage{nccmath}
\usepackage{caption}

\arxivnumber{} 

\title{\boldmath Particle and Superparticle Confinement in Higher Codimension Braneworlds.}

\author{F. E. A. de Souza, M. O. Tahim, R. I. de Oliveira Júnior, I. M. Macêdo}
\affiliation{Universidade Estadual do Ceará, Faculdade de Educação, Ciências e Letras do Sertão Central – R. Epitácio Pessoa, 2554, 63.900-000 Quixadá, Ceará, Brasil.\\
}

\emailAdd{francisco.emmanoel@uece.br}

\abstract{In this work we analyze the classical confinement of relativistic and supersymmetric spinning particles in higher-codimension braneworlds. Considering warped backgrounds generated by string-like $ n=2$ and global scalar defects $ n \geq 3$, we derive the effective radial dynamics from a Polyakov-type action. For spinless particles, the effective potential is monotonically decreasing, leading to repulsive behavior and the absence of confinement. In contrast, for $N=1,2 $ spinning particles, spin-curvature coupling modifies the potential, allowing the emergence of stable equilibrium points. Depending on the coupling parameter, particles may be confined on the membrane or in nearby regions, exhibiting bounded or satellite-like motion. These results emphasize the role of spin in localization mechanisms.}

\begin{document}
\maketitle
\flushbottom
\section{Introduction}

An essential ingredient in physics of extra dimensions is the decision related to the number of extra dimensions regarded in the model adopted. More specifically, in the braneworld context, where our physics is $4D$ and we have some ``freedom'' due to string theory bounds \cite{Polchinski:1998rq,Polchinski:1998rr}, we have several codimensions to study. In more realistic setups, solitonic solutions form the basic braneworld arena studied: if we discuss $5D$ physics, we walk on along the lines of Randall-Sundrum scenario with codimension $1$ membranes, $1D$  static kinks embedded in $5$ dimensions \cite{Liu:2017gcn}; If we go to $6D$ physics, then we use string like objects, codimension $2$ embedded into $6$ dimensions \cite{Gherghetta:2000qi, Oda:2000zc}; If we go one step further to $7D$, then we use codimension $3$ monopole like objects embedded in $7$ dimensions \cite{Gherghetta:2000jf, Olasagasti:2000gx}. This last one can be generalized to more dimensions if we built the codimension $n$ model. With this at hand, it is discussed the properties of $4D$ physics in each case cited, mainly those related to field localization. After all, we need to know if they are able to minimally retain the particle content of the standard model. In this sense, there are localization mechanisms that ensure the presence of fields of all spins needed \cite{Randall:1999vf,Randall:1999ee,Mukhopadhyaya:2007jn,Germani:2004jf,Alencar:2010mi,Tahim:2008ka,Fu:2016vaj,Lu:2024gmx,Guo:2023mki,Wan:2023usr}.

In this work we study the problems of relativistic particle and superparticle confinement in braneworlds. 
The relativistic particle is a good starting point to understand several aspects of physics. If we want to learn basic aspects of string theory we rely on it to make generalizations and propose, for instance, pure spinors as new variables \cite{Berkovits:2002zk,Berkovits:2001rb}. In particular, by studying extended supersymmetry models for the relativistic particle it is possible to find, after quantization, interesting representations of the usual field models \cite{Brink:1976uf, Gershun:1979fb, Howe:1989vn, Howe:1988ft, Siegel:1988ru}. This last point makes connection to the usual discussion of field localization. In fact, there are works where the issue of confinement of pure particle models gives the opposite answer: there is no natural particle confinement to a braneworld, unless the dilaton coupling is properly addressed. Despite this case, in particular, for the case of the $N=1$ spinning particle model in $D=5$, the particle is confined in another region, different from the membrane's position, such a characteristic which renders the non-confinement of the superparticle \cite{Souza:2019jqz}. In order to add the spin degrees of freedom at the classical level, the idea is to introduce Grassmann algebra in quantum mechanics \cite{Berezin:1976eg, Casalbuoni:1975hx, Casalbuoni:1975bj}. This was employed to represent a spinning point particle in flat spacetime \cite{Brink:1976uf}, which is described in terms of its position $x^{\mu}(\tau)$ and of an additional spin degree of freedom $\psi_{k}^{\mu}(\tau)$, an odd element of a Grassmann algebra with $\tau$ any parameter along the world line of the particle. In the work \cite{deSouza:2025wzv}, the confinement of $N=2$ superparticle, known to describe $p$- form fields after quantization, is discussed: as a result, the superparticle is not confined around the brane but, interestingly, has a satellite like behavior, being confined in another region. All of these results are related to codimension $1$ models, in other words, $5D$ with static kinks representing our $4D$ universe. The main question we would like to address in this present work is the following: Do higher codimension braneworlds have some impact in the problem of particle confinement? The answer we present is now positive.

This paper is organized in the following way. In section~\ref{sec:cod_n_problem}, we go directly to the Codimension $n$ problem, specializing in the Codimension $2$ and $n \geq 3$ cases. In section~\ref{sec:cod_spin_particle}, we discuss the problem of $N=2$ superparticle in the Codimension $n$ case. Finally, in section~\ref{sec:conclusion_perspective} we present our conclusions and perspectives.

\section{Codimension \texorpdfstring{$n$}{n} Problem} \label{sec:cod_n_problem}

In order to treat the general case for arbitrary codimension $n$, we first divide it into two cases: the $n=2$ case \cite{Gherghetta:2000qi, Oda:2000zc}, where we have a singular string-like defect generating a membrane, and the $n\geq3$ case \cite{Gherghetta:2000jf}, where a global topological defect is generated by a bulk scalar field. 

Then, we will analyze the behavior of the free particle in a region near the membrane ($ r =0$) in order to verify if there is some type of confinement.

\subsection{Codimension \texorpdfstring{$n=2$}{n=2}: the local string-like topological defect}
\label{sec:cod2}

For this case, we will use the metric solution for codimension $2$ due to a singular string-like defect to generate a membrane model. Its line element is given by \cite{Gherghetta:2000qi,Oda:2000zc}:
\begin{equation}\label{metriccd2}
ds^{2}=\sigma( r )\eta_{\mu\nu}dx^{\mu}dx^{\nu}+d r ^{2}+\gamma( r )d\theta^{2}, 
\end{equation}
where $0\leq  r <\infty$ is a radial extra dimension and $0\leq\theta<2\pi$ is an angular extra dimension. The nonzero components of the stress-energy tensor $T^B_A$ are assumed to be
\begin{equation}
    T^\mu_\nu=\delta^\mu_\nu f_0( r )\,,\,\,\, T^ r _ r =f_ r ( r )\,,\,\,\, T^\theta_\theta=f_\theta( r )\,,
\end{equation}
where we have introduced three source functions $f_0$ , $f_ r $ and $f_\theta$, which depend only on the radial
coordinate $ r $. They are interrelated by the equation
\begin{eqnarray}
    f^\prime_ r =2\frac{\sigma^\prime}{\sigma}(f_0-f_ r )+\frac12\frac{\gamma^\prime}{\gamma}(f_\theta-f_ r ).
\end{eqnarray}
We restrict this to the case where the four-dimensional cosmological constant is $\Lambda_{\text{phys}} = 0$, where it
is obtained a solution outside the core ($ r >\epsilon$) of the form 
\begin{equation}\label{sign2}
    \sigma( r )=e^{-c r },
\end{equation}
such that $\lim\limits_{ r \to 0}\sigma( r )=1$, and $\gamma( r )=R_{0}^{2}e^{-c r }$, with $R_0$ an arbitrary length scale. The solutions of the Einstein equations give us \cite{Gherghetta:2000qi}
\begin{equation}
    c=\sqrt{\frac25\frac{(-\Lambda)}{M^{4}_6}},  \label{c2}
\end{equation}
and the negative exponential solution \eqref{sign2} requires that $\Lambda<0$. Therefore, the non-null connections are given by
\begin{eqnarray}
    &&\Gamma^{\mu}_{ r \nu}=\frac{1}{2}\frac{\sigma( r )'}{\sigma( r )}\delta^{\mu}_{\nu}\label{gammax}\\
    &&\Gamma^{ r }_{\theta\theta}=-\frac{1}{2}\gamma( r )'\,,\,\,\,\,\Gamma^{ r }_{\mu\nu}=-\frac12 \sigma( r )'\eta_{\mu\nu}\label{gammaw}\\
    &&\Gamma^{\theta}_{ r \theta}=\frac{1}{2}\frac{\gamma( r )'}{\gamma( r )}\label{gammay}
    \end{eqnarray}

The next step is to understand the motion of a particle in this background. We consider the action to the test particle to be Polyakov like and given by
\begin{equation}\label{actionE}
    S=\frac{1}{2}\int\left[\frac{1}{e}g_{AB}\dot{x}^{A}\dot{x}^{B}-em^{2}\right]d\tau .
\end{equation}
This is the action that we will work on in the next sections. From this we obtain the equations of motion
\begin{eqnarray}
    &&\frac{\delta S}{\delta e}\rightarrow g_{AB}\dot{x}^{A}\dot{x}^{B}+e^{2}m^{2}=0\label{motE}\\
    && \frac{\delta S}{\delta \dot{x}^{C}}\rightarrow \frac{D}{D\tau}[e^{-1}g_{AC}\dot{x}^{A}]=0.\label{motX}
\end{eqnarray}
where the equation to $e$ is not of a dynamical variable.

Then, using the Christoffel Symbols \eqref{gammax}, \eqref{gammaw}, \eqref{gammay} and fixing the gauge as $e=1$ in \eqref{motE}, we obtain the equations to $x^{A}$ as follows:
\begin{eqnarray}
    &&\frac{D}{D\tau}[\dot{x}^{\mu}]=0\rightarrow\ddot{x}^{\mu}+\sigma( r )^{-1}\sigma( r )'\delta^{\mu}_{\nu}\dot{x}^{\nu}\dot{ r }=0 \label{mot1} \\
    &&\frac{D}{D\tau}[\dot{ r }]=0\rightarrow\ddot{ r }-\frac{1}{2}[\eta_{\mu\nu}\dot{x}^{\mu}\dot{x}^{\nu}+R_{0}^{2}\dot{\theta}^{2}]\sigma( r )'=0\label{mot2}\\
    &&\frac{D}{D\tau}[\dot{\theta}]=0\rightarrow\ddot{\theta}+\gamma( r )^{-1}\gamma( r )'\dot{ r }\dot{\theta}=0.\label{mot3}
\end{eqnarray}

In order to obtain the conserved quantities, we need to work with the expressions above together and find total derivatives. Multiplying \eqref{mot1} by $\sigma( r )$ we can rewrite it as:
\begin{equation}\label{momentxn=2}
    \frac{d}{d\tau}[\sigma( r )\dot{x}^{\mu}]=0\rightarrow p^{\mu}=\sigma( r )\dot{x}^{\mu},
\end{equation}
where $p^{\mu}$ is the conserved momentum on the hypersurfaces along the extended extra-dimension $r$. For the spherical extra-dimension $\theta$, we find
\begin{equation}
    \frac{d}{d\tau}[{\gamma( r )}\dot{\theta}]=0\rightarrow{C}_{\theta}={\gamma( r )}\dot{\theta},\label{momentthetan=2}
\end{equation}
which give the conserved quantity in the spherical extra-dimension. Finally, by using the constraint \eqref{motE} with the gauge $e=1$ fixed previously, we can rewrite \eqref{mot2} as:
\begin{equation}
  \frac{d}{d\tau}\left[\frac{1}{2}(\dot{ r }^{2}+m^{2})\sigma( r )\right]=0\rightarrow C( r )=\frac{1}{2}(\dot{ r }^{2}+m^{2})\sigma( r ),\label{momentwn=2}
\end{equation}
where $C( r )$ is a constant. Therefore, we can take $C( r )=C( r )|_{ r =0}$ and, following \cite{Dahia:2007ep}, we construct an expression for an effective potential  acting on the particle along the radial extra-dimension obtaining
\begin{equation}    \sigma( r )\dot{ r }^{2}=\dot{ r }_{0}^{2}-m^{2}[\sigma( r )-1].
\end{equation}
This last equation tells us that the particle has $\dot{ r }_{0}^{2}$ as the amount of initial energy. Then the effective potential acts on the particle such that the inequality $\sigma( r )\dot{ r }^{2}\leq \dot{ r }_{0}^{2}$ is satisfied with it given by 
\begin{equation}\label{ueffcd2}
  U_{eff}( r )=m^{2}[\sigma( r )-1],
\end{equation}
where we define an effective potential density by $u_{eff}( r )={U_{eff}( r )}/{m^2}$ as:
\begin{equation}\label{ueffcd2dens}
   u_{eff}( r )=\sigma( r )-1
\end{equation}
We can note that the conserved quantities obtained for $x^{\mu}$ and $ r $ are given, respectively, by \eqref{momentxn=2} and \eqref{momentwn=2}. And by symmetry, they are independent of the spherical extra-dimensions, just as the effective potential above. 

\begin{figure}[!ht]
    \centering
    \includegraphics[width=0.8\textwidth]
    {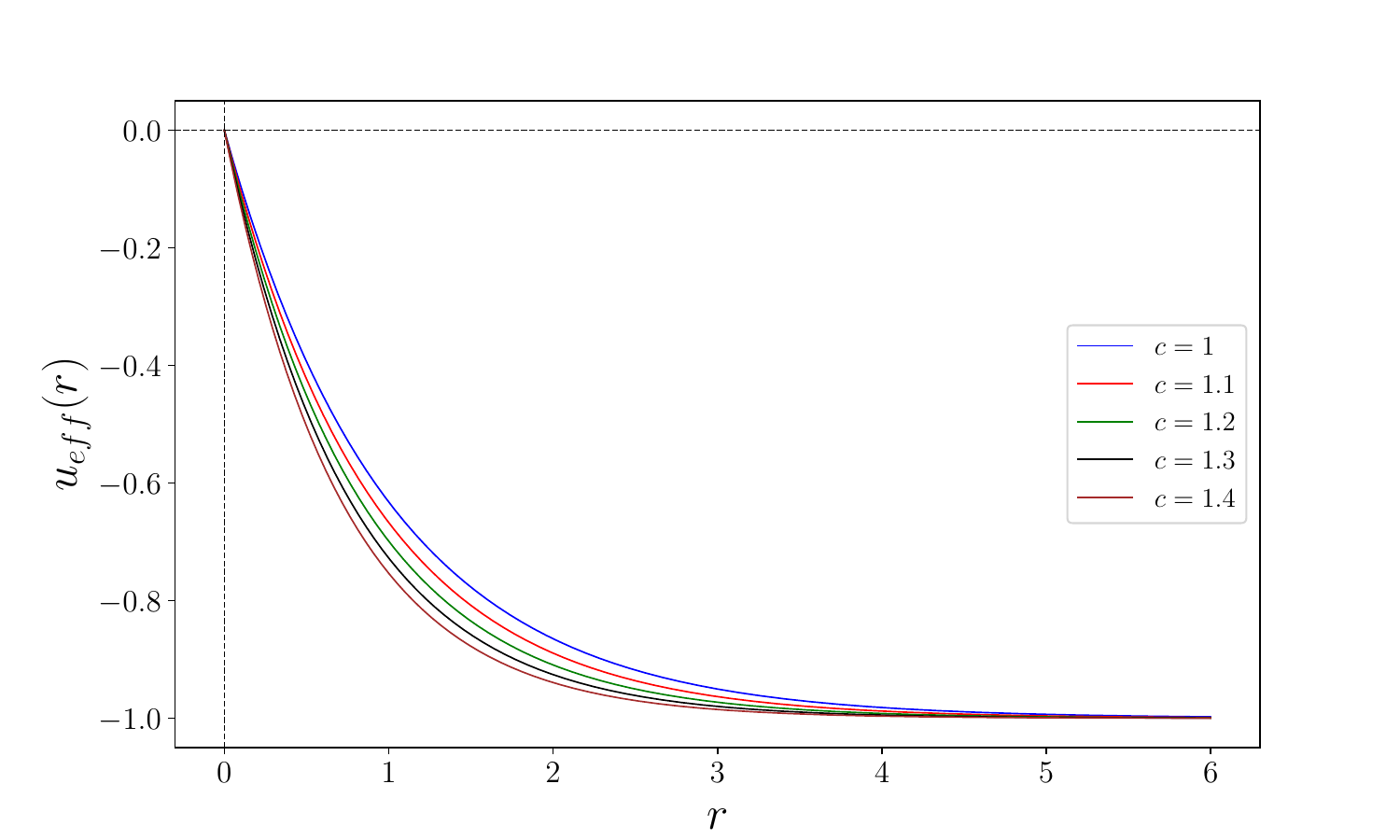}
     \caption{The effective potential for a spinless massive particle in codimension 2.}
    \label{fig:Pot_HedgehogNOspin}
\end{figure}

We now analyze the behavior of a spinless particle along the radial extra dimension, Figure \ref{fig:Pot_HedgehogNOspin}. We consider the effective potential \eqref{ueffcd2dens} for $\sigma$ given by \eqref{sign2} and the constant $c$ by \eqref{c2}. We set $c$ [1, 1.1, 1.2, 1.3, 1.4 ] to reproduce the background. Effectively, there is no confinement.
Expanding the potential near the origin gives
\begin{equation}
u_{eff} \approx -cr + \frac{c^2}{2}r^2 + \dots
\end{equation}
The particle is subject to the effective force density $f_{eff} = -u_{eff}'$. The force on the particle at the origin  is
\begin{equation}
f_{eff}(0) = c > 0,
\end{equation}
and then it is repelled from there. For arbitrary values of $r$ the effective force is given by
\begin{equation}
f_{eff} = c e^{-cr} > 0, \quad \forall  r > 0.
\end{equation}
Hence, the effective potential is strictly decreasing from $u_{eff}(0) = 0$ to
\begin{equation}
u_\infty = \lim_{r \rightarrow \infty} u_{eff}(r) = -1.
\end{equation}
Therefore, there are no critical points for $r > 0$, and the origin is not an equilibrium point since the force is nonzero. Consequently, there is no stable equilibrium anywhere in the system. For a classical particle, the force is always positive and points outward. A particle placed at any finite radius will accelerate toward larger $r$. If the particle comes from infinity with total energy $E > -1$, it will be reflected at the point where $u_{eff} = E$ and return to infinity — a scattering trajectory. The conclusion is that confinement does not occur.

\newpage
\subsection{Codimension \texorpdfstring{$n\geq3$}{n>=3}: the bulk scalar field as topological defect}

\begin{figure}[!ht]
    \centering
    \includegraphics[width=0.8\textwidth]
    {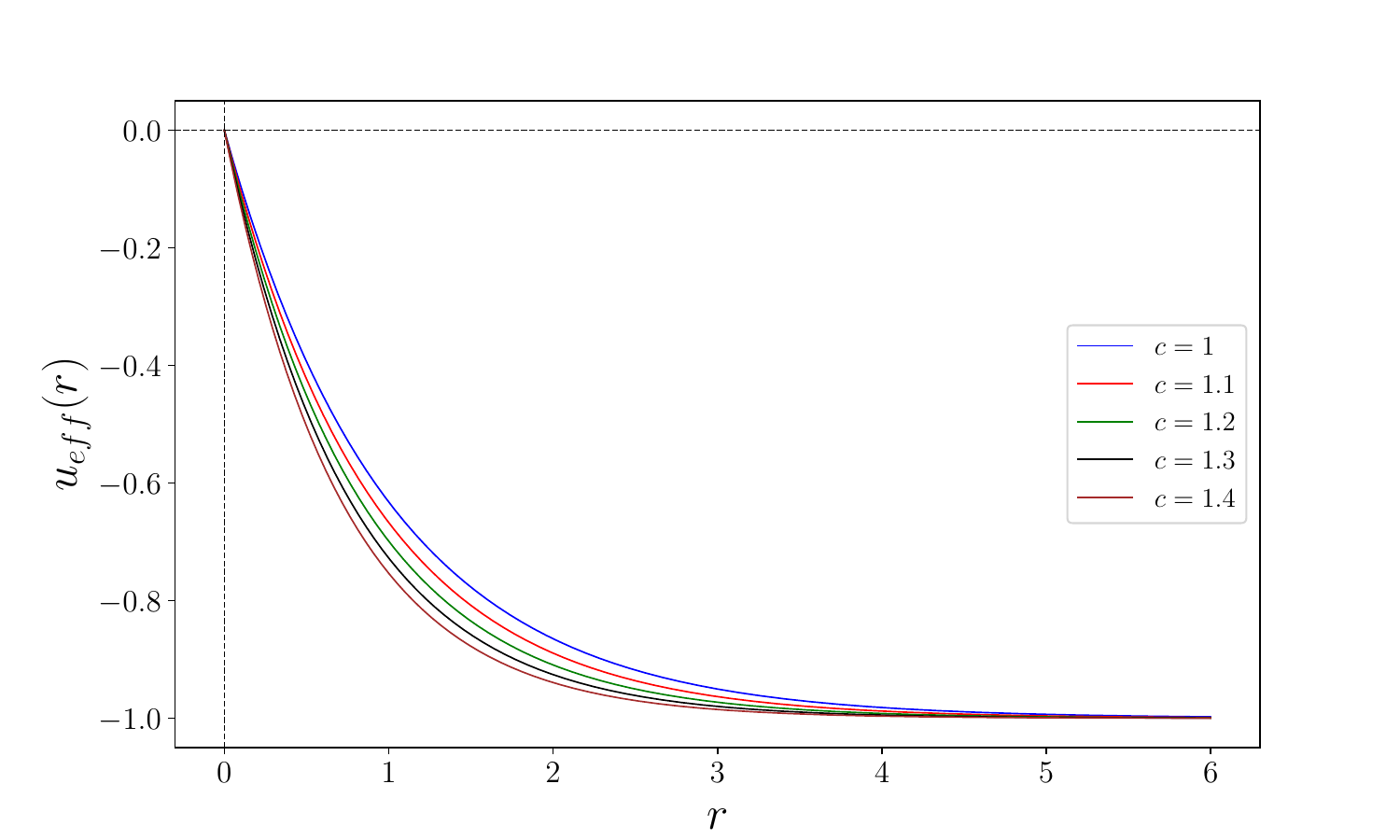}
     \caption{The effective potential for a spinless massive particle in codimension $n\geq3$.}
    \label{fig:Pot_HedgehogNOspin3}
\end{figure}

\begin{figure}[!ht]
    \centering
    \includegraphics[width=0.8\textwidth]
    {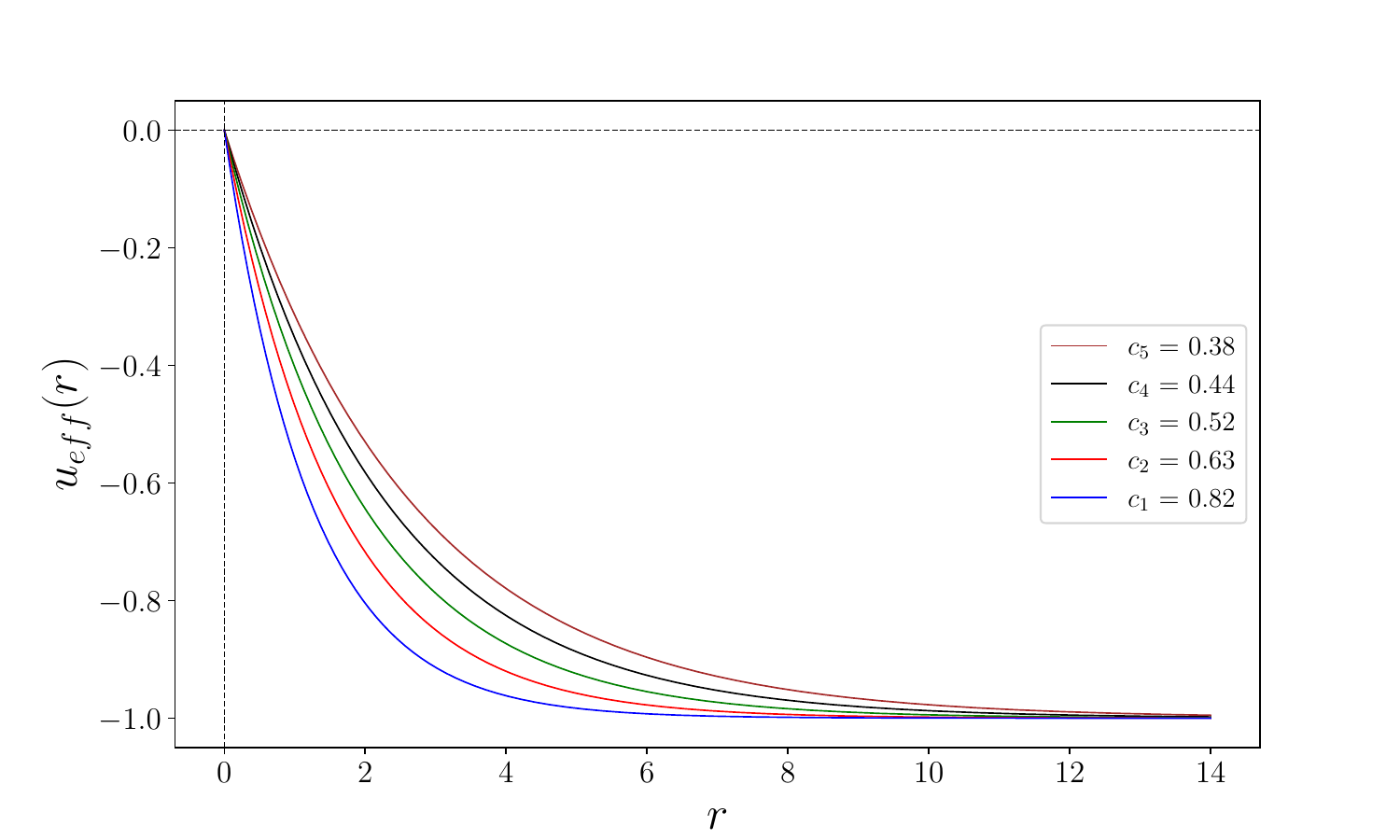}
     \caption{The effective potential for a spinless massive particle in codimension $n\geq3$. We set $c$ with $n$ contributions.}
    \label{fig:Pot_HedgehogNOspin3_C}
\end{figure}

In this case, we consider a \(D\)-dimensional spacetime \cite{Olasagasti:2000gx, Gherghetta:2000jf}. We adopt the notation \(\{x^\mu\}\), with \(\mu = 0, 1, \dots, p-1\), for coordinates on the brane worldsheet; for unit vectors in the extra dimensions we use \(\{d^a\}\), with \(a = 1, 2, \dots, n\); for the general coordinates of the \(D\)-dimensional spacetime, where \(D = p + n\), we use \(\{x^A\}\), with \(A = 0, 1, \dots, D-1\). For simplicity, we are interested in the case where the four-dimensional cosmological constant is given by $\Lambda_{\text{phys}}=0$. This global defect in $n$ extra dimensions is described by the Lagrangian \cite{Olasagasti:2000gx}:
\begin{equation}
    L=\frac12\, \partial_A\phi^a \, \partial^A\phi^a-V(\phi),
\end{equation}
where a global defect in $n$ extra dimensions is described by a multiplet of $n$ scalar fields $\{\phi^{a}\}$ with a potential \cite{Gherghetta:2000jf}
\begin{equation}
    V(\phi)=\lambda(\phi^{a}\phi^{a}-v^2)^2.
\end{equation}
This potential minimum is described by the $n$-sphere $\phi^{a}\phi^{a}=v^{2}$. The defect solution should have $\phi^{a}=0$ at the center of the defect and approach the radial `hedghog' configuration outside the core,
\begin{equation}\label{scalarhog}
    \phi^{a}( r )=vd^{a}
\end{equation}
with $d^{a}d^{a}\equiv 1$. The spherical coordinates in the extra dimensions are defined by the usual relations $d^{a}=\{\cos{\theta_1,\sin{\theta_1\cos{\theta_2,\dots}}}\}$  with $0\leq  r \leq\infty$, $0\leq\{\theta_{n-1},\dots,\theta_2\}\leq\pi$ and $0\leq\theta_1\leq2\pi$. For the exterior solutions $V(\phi)\approx0$ and $\phi^{a}( r )$ is accurately approximated by \eqref{scalarhog} \cite{Olasagasti:2000gx, Gherghetta:2000jf}. The scalar field gives an additional contribution to the stress-energy tensor in the bulk with components
\begin{equation}
    T^{\mu}_\nu=(n-1)\frac{v^2}{2\gamma}\delta^{\mu}_\nu\,;\,\,     T^{ r }_ r =(n-1)\frac{v^2}{2\gamma}\,;\,\, T^{\theta}_\theta=(n-3)\frac{v^2}{2\gamma}\,.
\end{equation}
Therefore, the spacetime is described by the metric \cite{Gherghetta:2000jf}
\begin{equation} \label{metricgheta}
ds^{2}=\sigma( r )\eta_{\mu\nu}dx^{\mu}dx^{\nu}+d r ^{2}+\gamma( r )d\Omega_{n-1},
\end{equation}
with 
\begin{equation}
    \sigma( r )=e^{-c r }.\label{sigman3+}
\end{equation}
When $v^2=(n-2)M^{n+2}_D$, the Einstein's equations give us:
\begin{eqnarray}
&&\gamma( r )=R_{0}^{2}\sigma( r );\\
&&c=\sqrt{\frac{8(- \Lambda_D)}{(n+2)(n+3)M^{n+2}_D}}\,, \label{cn3}
\end{eqnarray}
where $R_0$ is an arbitrary length scale. In this case gravity is localized on the membrane for $n\geq3$ if we have a global topological defect generated by a bulk scalar field \cite{Gherghetta:2000jf}. Recursively, $d\Omega^{2}_{n-1}$ is defined as \cite{Olasagasti:2000gx, Gherghetta:2000jf}
\begin{equation}
    d\Omega^{2}_{n-1}=d\theta_{n-1}^{2}+\sin^{2}\theta_{n-1}d\Omega^{2}_{n-2},
\end{equation}
and stands for the metric on a unit $(n-1)$-sphere that we will be represented by the ansatz
\begin{equation}
    d\Omega^{2}_{n-1}\equiv\xi_{ij}(\theta)d\theta_{i}d\theta_{j},
\end{equation}
where $\xi_{ij}$ is an Euclidean diagonal matrix that is a function of the spherical extra dimensions $\theta_i$. Therefore, we can rewrite the metric \eqref{metricgheta} as:
\begin{equation} \label{metriccdn}   ds^{2}=\sigma( r )\eta_{\mu\nu}dx^{\mu}dx^{\nu}+d r ^{2}+\gamma( r )\xi_{ij}(\theta)d\theta_{i}d\theta_{j}
\end{equation}
where  this metric is given by the diagonal matrix with $(p+n)$-dimension 
\begin{equation}
g_{AB}=\text{diag}[\sigma( r )\eta_{\mu\nu},1,\gamma( r )\xi_{ij}]
\end{equation}
with $\mu,\nu=0,1,...p-1$ and $i,j=1,2,\dots,n-1$. The signature of the space-time is $\eta_{\mu\nu}=\text{diag}(-1,+1,\dots,+1)$ and $\xi_{ij}$ is the metric of the internal space. Then, the non-null connections are given by
\begin{eqnarray}
    &&\Gamma^{\mu}_{ r \nu}=\frac{1}{2}\frac{\sigma'( r )}{\sigma( r )}\delta^{\mu}_{\nu};\label{gammast}\\
    &&\Gamma^{ r }_{ij}=-\frac{1}{2}\gamma'( r )\xi_{ij},\,\,\Gamma^{ r }_{\mu\nu}=-\frac{1}{2}\sigma'( r )\eta_{\mu\nu};\label{gammanw}\\    &&\Gamma^{i}_{jk}=\tilde{\Gamma}^{i}_{jk},\,\,\Gamma^{i}_{ r  k}=\frac{1}{2}\frac{\gamma'( r )}{\gamma( r )}\delta^{i}_{k},\label{gammaij}
    \end{eqnarray}
where $$\tilde{\Gamma}^{i}_{jk}=\frac12 \xi^{il}(\partial_{j}\xi_{kl}+\partial_{k}\xi_{jl}-\partial_{l}\xi_{jk}).$$
Taking the action \eqref{actionE}, we obtain the equations of motion 
\eqref{motE} and \eqref{motX}, such that, using the Christoffel Symbols \eqref{gammast}, \eqref{gammanw}, \eqref{gammaij} and fixing the gauge $e=1$, we obtain the constraint equation
\begin{equation}    \sigma( r )\eta_{\mu\nu}\dot{x}^{\mu}\dot{x}^{\nu}+\dot{ r }^2+\gamma( r )\xi_{ij}\dot{\theta}^{i}\dot{\theta}^{j}+m^2=0.\label{constcodm3}
\end{equation}
By equation \eqref{motE} and equation \eqref{motX}, we obtain the equations to $x^{A}$ as follows:

\begin{eqnarray}
    &&\frac{D}{D\tau}[\dot{x}^{\mu}]=0\rightarrow\ddot{x}^{\mu}+\frac{\sigma'( r )}{\sigma( r )}\delta^{\mu}_{\nu}\dot{x}^{\nu}\dot{ r }=0 \label{mot1gen} \\
    &&\frac{D}{D\tau}[\dot{ r }]=0\rightarrow\ddot{ r }-\frac{1}{2}[\eta_{\mu\nu}\dot{x}^{\mu}\dot{x}^{\nu}+R_{0}^{2}\xi_{ij}\dot{\theta_{i}}\dot{\theta_{j}}]\sigma'( r )=0.\label{mot2gen}\\
    &&\frac{D}{D\tau}[\dot{\theta}^{i}]=0\rightarrow\ddot{\theta}^i+\tilde{\Gamma}^{i}_{jk}\dot{\theta}^{j}\dot{\theta}^{k}+\frac{\gamma'( r )}{\gamma( r )}\dot{\theta}^{i}\dot{ r }=0\label{mot3gen}  .  
\end{eqnarray}
In order to obtain the conserved quantities, we need again to work with the expressions above to find total derivatives. Multiplying \eqref{mot1gen} by $\sigma( r )$ we can rewrite it as:
\begin{equation}\label{momentxn}
    \frac{d}{d\tau}[\sigma( r )\dot{x}^{\mu}]=0\rightarrow p^{\mu}=\sigma( r )\dot{x}^{\mu},
\end{equation}
where $p^{\mu}$ is the conserved momentum in the hypersurfaces along the extended extra-dimension $ r $. The conserved quantities obtained from the equation \eqref{mot3gen}, depends on the form of the spherical affine connection $\tilde{\Gamma}^{i}_{jk}$. Finally, by using the constraint \eqref{motE}, we can rewrite \eqref{mot2gen} as:
\begin{equation}
  \frac{d}{d\tau}\left[\frac{1}{2}(\dot{ r }^{2}+m^{2})\sigma( r )\right]=0\rightarrow C( r )=\frac{1}{2}(\dot{ r }^{2}+m^{2})\sigma( r ),
\end{equation}
where $C( r )$ is a constant. Therefore we can take $C( r )=C( r )|_{ r =0}$ to construct an expression for an effective potential acting on the particle. We obtain
\begin{equation}
    \sigma( r )\dot{ r }^{2}=\dot{ r }_{0}^{2}-m^{2}[\sigma( r )-1].
\end{equation}
This last equation tells us that the particle has $\dot{ r }_{0}^{2}$ as the amount of initial energy. Then again the effective potential acts on the particle such that the inequality $\sigma( r )\dot{ r }^{2}\leq \dot{ r }_{0}^{2}$ is satisfied by 
\begin{equation}\label{ueffcd3+}
  u_{eff}( r )=\sigma( r )-1.  
\end{equation}
This effective potential is the same as that obtained for codimension $2$, except for a constant $c$ given by \eqref{cn3}. The Figure~\ref{fig:Pot_HedgehogNOspin3} also assumes the background $c$ [1, 1.1, 1.2, 1.3, 1.4 ] and, for that reason, reproduce the same behavior. On the other hand, if we adjust the $c$ parameter according to the choice of $n$, keeping the choice of parameter $\Lambda_{D} = -1$ and Mass $M_{D} = 1$ fixed as shown in the Figure~\ref{fig:Pot_HedgehogNOspin3_C}, we see a more pronounced behavior. Consequently, the same conclusions hold as in Sec. \ref{sec:cod2} in the same background: the particle experiences a repulsive force for all $r > 0$, there are no equilibrium points, and confinement does not occur. The system supports only scattering trajectories

\newpage

\section{Codimension \texorpdfstring{$n$}{n}: The Spinning particles \texorpdfstring{$N=1,2$}{N=1,2}}\label{sec:cod_spin_particle}

In this section, we generalize the particle action in order to obtain extended supersymmetry $N=1, 2$ \cite{Berezin:1976eg, Brink:1976uf, Brink:1976sc, Brink:1976sz, Rietdijk:1989qa, Gershun:1979fb, Howe:1989vn, Howe:1988ft, Siegel:1988ru, Rivelles:1990dq} for a codimension $n$ topological defect. Following the idea proposed by the action \eqref{fullspinac}, the case $N=1$ must represent a description of the spin $\frac12$ particle \cite{Souza:2019jqz, Brink:1976uf}, and  the case $N=2$ must represent a description of the gauge field \cite{deSouza:2025wzv, Howe:1989vn}. It is well known that, in this case, the Grassmann variables contribute to the effective potential \cite{Souza:2019jqz, deSouza:2025wzv}. The question we pose is whether their contributions are relevant to the problem of particle confinement in this new case. We consider the following action:

\begin{eqnarray}\label{fullspinac}
    S&=&\int\Bigg[\frac{e^{-1}}{2}g_{AB}\dot{x}^{A}\dot{x}^{B}-\frac{e}{2}m^{2}-\frac{i}{2}g_{AB}\psi^{A}_{k}\frac{D\psi^{B}_{k}}{D\tau}-\frac{i}{2}\psi^{5}_{k}\dot{\psi^{5}_{k}}\Bigg.\nonumber\\
    &-&\Bigg.\frac{i}{2}\lambda^{k}(e^{-1}g_{AB}\psi^{A}_{k}\dot{x}^{B}+m\psi^{5}_{k})-\frac{1}{8}e^{-1}g_{AB}(\lambda_{k}\psi^{A}_{k})(\lambda_{k}\psi^{B}_{k})\Bigg.\nonumber\\    &+&\Bigg.f\left[-i\frac{1}{2}\epsilon_{kl}(g_{AB}\psi^{A}_{k}\psi^{B}_{l}+\psi^{5}_{k}\psi^{5}_{l})-\Big(q-\frac{1}{2}D\Big)\right]\Bigg]d\tau,
\end{eqnarray}
where $$\frac{D\psi^P}{D\tau} = \dot{\psi}^P + \Gamma^{P}_{QR} {\psi}^Q \dot{x}^R.$$ By Hamilton principle, we obtain from (\ref{fullspinac}) the following equations of motion:
\begin{eqnarray}
\frac{\delta S}{\delta e} = 0 &\to& g_{AB}\left[\dot{x}^{A}\dot{x}^{B}-i\lambda_{k}\dot{x}^{A}\psi^{B}_k-\frac{1}{4}(\lambda_{k}\psi^{A}_{k})(\lambda_{k}\psi^{B}_{k})\right]+e^{2}m^2 = 0,\label{402}\\
\frac{\delta S}{\delta \lambda_{k}} = 0 &\to & g_{AB}\left[\dot{x}^{A}\psi^{B}_k-\frac{i}{2}(\lambda_{k}\psi^{A}_{k})\psi^{B}_{k}\right]+em\psi^{5}_{k}=0,\label{404}\\
\frac{\delta S}{\delta f} = 0 &\to& i\frac{1}{2}\epsilon_{kl}(g_{AB}\psi^{A}_{k}\psi^{B}_{l}+\psi^{5}_{k}\psi^{5}_{l})+\Big(q-\frac{1}{2}D\Big)=0,\label{409}\\
\frac{\delta S}{\delta \psi^{B}_{k}} = 0 &\to& \frac{D\psi^{B}_{k}}{D\tau}-\frac{1}{2}e^{-1}\lambda_{k}\dot{x}^{B}+\frac{i}{4}e^{-1}\lambda_{k}(\lambda_{k}\psi^{B}_{k})+f\epsilon_{lk}\psi^{B}_{l}=0,\label{403}\\
\frac{\delta S}{\delta \psi^{5}_{k}}= 0 &\to & \dot{\psi^{5}_{k}}-\psi^{5}_{l}f\epsilon_{kl}-\frac{1}{2}\lambda_{k}m=0,\label{405}\\
\frac{\delta S}{\delta x^{A}} = 0 &\to& \frac{D}{D\tau}\left[e^{-1}g_{AF}\dot{x}^{F}\right]-\frac{i}{2}\frac{d}{d\tau}\left[\lambda_{k} e^{-1}g_{AF}\psi^{F}_{k}\right]\nonumber\\& &+\frac{i}{2}R_{ASQR}\psi^{Q}_{k}\psi^{R}_{k}\dot{x}^{S}=0.
\label{406}
\end{eqnarray}

For $N=2$, the worldline reparameterization, supersymmetry transformations and the local worldline $SO(2)$ transformations due to the introduction of the Chern-Simons term \cite{Howe:1988ft, Howe:1989vn, Rivelles:1990dq}, are satisfied if we choose the gauge conditions $ \dot \lambda_k = \dot e = \dot f = 0$ in order to determine the gauge parameters \cite{Rivelles:1990dq}. Thus, fixing the gauge as $\lambda_k = -\Big(\frac2m f\Big)\epsilon_{kl} \psi^5_l$ and $e=f=1$, we simplify equations (\ref{402})-(\ref{406}), corresponding to the orthonormal gauge \cite{Brink:1976sc}, to the following equations
\begin{eqnarray}
&&\frac{D\psi^{P}_{k}}{D\tau}=0,\label{503}\\
&&\dot{\psi^{5}_{k}}=0,\\
&&\frac{D\dot{x}^{N}}{D\tau}+\frac i2 R^{N}_{\ SQR}\psi^{Q}\psi^{R}\dot{x}^{S}=0.\label{506}
\end{eqnarray}
and constraints
\begin{eqnarray}
&&g_{PQ}\dot{x}^{P}\dot{x}^{Q}+m^{2}=0,\label{con1}\\
&&g_{PQ}\dot{x}^{P}\psi^{Q}_{k}+m\psi^{5}_{k}=0,\label{con2}\\
&&i\epsilon_{kl}(g_{AB}\psi^{A}_{k}\psi^{B}_{l}+\psi^{5}_{k}\psi^{5}_{l})=0\label{con3}.
\end{eqnarray}
We are using units $\hslash = 1 $. By restoring $\hslash$, the Chern-Simons term $ -f(q-\frac d2)$ is multiplied by $\hslash$ and therefore it vanishes in the classical limit \cite{Howe:1989vn}. The action for $N = 1$ does not contain the Chern-Simons term. However, the same equations of motion and constraints are obtained \cite{Gershun:1979fb, Berezin:1976eg, Brink:1976sz, Brink:1976uf, Rietdijk:1989qa}.

In order to study the general case for arbitrary codimension $n$, we will consider the cases of a local string-like topological defect with codimension $n=2$ and a bulk scalar field as a topological defect with codimension $n\geq3$. Then, we can analyze the behavior of the spinning particle $N=1, 2$ in codimension $n$. In this sense, we need to obtain an effective potential to describe, as in \cite{deSouza:2025wzv}, the behavior of the particle in the region near the membrane.

\subsection{Spinning Particle in Codimension \texorpdfstring{$n=2$}{n=2}: the local string-like topological defect}\label{spincd2}

We will adopt the metric \eqref{metriccd2} to describe the background in codimension $n=2$. Then, the Christoffel symbols are given by \eqref{gammax}, \eqref{gammaw} and\eqref{gammay}. Taking the equation \eqref{503} for $\mu$, $ r $ and $\theta$ we obtain
\begin{eqnarray}
    &&\dot{\psi\,^\mu _k} + \frac12 \frac{\sigma'}{\sigma}(\psi_k ^ r  \dot{x}^\mu + \psi_k ^\mu \dot{ r }) = 0 \label{psiu},\\
     &&\dot{\psi^ r  _k} + \frac12 \frac{\sigma'}{\sigma}(\psi_k ^ r  \dot{ r } + m \psi_k ^5) = 0 \label{psip},\\
     &&\dot{\psi}^\theta +\frac12 \frac{\sigma'}{\sigma}(\dot{r}\psi^\theta + \dot{\theta}\psi^r) = 0.\label{psiteta}
\end{eqnarray}
where we use the constraint \eqref{con2} in \eqref{psip}. Then, we can rewrite \eqref{psip} and find the solution:
\begin{equation}
    \psi^ r _k(\tau)\psi^5 _k = \psi^ r _k(0)\psi^5 _k \sigma^{-\frac12},\label{constp5}
\end{equation}
where $\gamma' = \gamma\frac{\sigma'}{\sigma}$ and $\psi^5 _k\psi^5 _k = 0$. Now, we can multiply \eqref{psiu} and \eqref{psip} by $\psi^ r  _k$ and $\psi^\mu _k$, respectively, to get
\begin{equation}
   \frac{d}{d\tau}(\sigma \psi^ r _k \psi^\mu_k) + \frac12 m\sigma' \psi^5_k \psi^\mu_k = 0, \label{psipu}
\end{equation}
and multiply \eqref{psip} and \eqref{psiteta} by $\psi^5_k \psi^\theta_k$  and $\psi^5_k \psi^r_k$ respectively to get the conserved quantity
\begin{equation}
    \frac{d}{d\tau}(\sigma \psi^5 \psi^r \psi^\theta) = 0.\label{psitetacon}
\end{equation}

Now, to obtain the equations of motion for the $x^P$ direction in \eqref{506}, we need to obtain the Riemann tensor components, which are given by:
\begin{eqnarray}
    &&R^{\mu}_{\nu\kappa\tau}=\frac{1}{4}\frac{\sigma'^{2}( r )}{\sigma^2(r)}\Big[\delta^{\mu}_{\tau}g_{\nu\kappa}-\delta^{\mu}_{\kappa}g_{\nu\tau}\Big];\label{ric1}\\
    &&R^{\mu}_{rr\nu}=\left[\frac{1}{2}\frac{\sigma''( r )}{\sigma( r )}-\frac{1}{4}\frac{\sigma'^{2}( r )}{\sigma^{2}( r )}\right]\delta^{\mu}_{\nu};\label{ric2}\\
    &&R^{\mu}_{\theta\theta\nu}=\frac{1}{4}\frac{\sigma'^2( r )}{\sigma^2( r )}\gamma(r)\delta^{\mu}_{\nu};\label{ric3}\\
    &&R^{ r }_{\mu\nu  r }=\left[\frac{1}{2}\frac{\sigma''( r )}{\sigma( r )}-\frac{1}{4}\frac{\sigma'^{2}( r )}{\sigma^2( r )}\right]g_{\mu\nu};\label{ric4}\\
    &&R^{ r }_{\theta\theta  r }=\left[\frac{1}{2}\gamma''( r )-\frac{1}{4}\frac{\gamma'^{2}( r )}{\gamma( r )}\right];\label{ric5}\\
    &&R^{\theta}_{ r  r \theta}=\left[\frac{1}{2}\frac{\gamma''( r )}{\gamma( r )}-\frac{1}{4}\frac{\gamma'^{2}( r )}{\gamma^{2}( r )}\right].\label{ric6}
\end{eqnarray}
Thus, using \eqref{mot1}-\eqref{mot3}, \eqref{constp5}, \eqref{psipu}, \eqref{psitetacon}, \eqref{ric1}-\eqref{ric6} in \eqref{506}, we get the  conserved quantities in the directions $x^\mu$ , $r$ and $ \theta$ given by:
\begin{eqnarray}
 &&\frac{d}{d\tau}(\sigma \dot{x}^\mu + \frac i2 \psi^r_k \psi^\mu_k\sigma') = 0;\label{consx} \\
 &&\frac{d}{d\tau}\Bigg[\sigma \dot{\theta}\psi^5_k + \frac i2 \psi^5 \psi^r(0)\psi^\theta\sigma'\sigma^{-\frac12}\Bigg] = 0;\label{consteta}\\
 && \frac{d}{d\tau}[\sigma(\dot{r}^2 + m^2)-im\psi^5_k \psi^r_k(0)\sigma'\sigma^{-\frac12}] = 0 \label{consr}. 
\end{eqnarray}
Here we define the conserved momentum
\begin{equation}
    p^\mu = \sigma \dot{x}^\mu + \frac i2 \psi^r_k \psi^\mu_k\sigma'.\label{consmomentum}
\end{equation}
The conserved angular quantity \eqref{consteta},
\begin{equation}
\frac{d}{d\tau}\Bigg[\sigma \dot{\theta}\psi^5_k + \frac i2 \psi^5 \psi^r(0)\psi^\theta\sigma'\sigma^{-\frac12}\Bigg] = 0 
\end{equation}
acts as a spin angular momentum. The first term, $\sigma \dot{\theta}\psi^5_k$, couples the conserved angular momentum \eqref{momentthetan=2} with the fermionic chirality operator, since after quantization, $\psi^5$ corresponds to the Dirac matrix $\gamma^5$ as the chirality operator \cite{Brink:1976uf, Gershun:1979fb, Howe:1988ft, Howe:1989vn}. The spin connection term, $\frac i2 \psi^5 \psi^r(0)\psi^\theta\sigma'\sigma^{-\frac12}$, acts to preserve chirality in a codimension-2 background with radial curvature. This contribution functions as a correction to the Fermi-Walker transport. By equation \eqref{consr}, we have the conserved quantity
\begin{equation}
    Q(r) = \sigma(\dot{r}^2 + m^2)-im\psi^5_k \psi^r_k(0)\sigma'\sigma^{-\frac12},
\end{equation}
which is a constant. Then, if we compare $Q(r) = Q(r)|_{r=0}$ we get
\begin{eqnarray}
   &&\sigma(\dot{r}^2 + m^2)-im\psi^5_k \psi^r_k(0)\sigma'\sigma^{-\frac12} = \dot{r_0}^2 + m^2-im\psi^5_k \psi^r_k(0)\sigma'(0)\nonumber\\
   &&\sigma \dot{r}^2 = \dot{r_0}^2 - m^2\Bigg[ \sigma - \frac im \psi^5_k \psi^r_k(0)\Big(\sigma'\sigma^{-\frac12}-\sigma'(0)\Big) -1 \Bigg]\label{pot}.
\end{eqnarray}
This last equation tells us that the particle has $\dot{ r }_{0}^{2}$ as the amount of initial energy along the radial extra dimension. Then, an effective potential acts on the particle such that the inequality $\sigma( r )\dot{ r }^{2}\leq \dot{ r }_{0}^{2}$ is satisfied by the effective potential density 
\begin{equation}\label{effn2}
u_{eff} =  \sigma - A\Big[\sigma'\sigma^{-\frac12}-\sigma'(0)\Big] -1.
\end{equation}
When $\psi^N_k$ is zero, $A=0$ and we recover the results discussed previously for a bosonic test particle given by equations \eqref{ueffcd2dens} and \eqref{ueffcd3+}. 

We obtain, due to the spin variable, a new piece in $u_{eff}$ acting on the particle: the spin modifies the effective potential through its interaction with the curvature. This opens up the possibility of confining a spinning particle \cite{Souza:2019jqz}.
We proceed to check that now. For $\sigma(r)$ given by \eqref{sign2} and near the origin the potential expands as 
\begin{equation}
    u_{eff}(r) \approx \Bigg(-c - \frac{Ac^2}{2} \Bigg)r + \frac{c^2}{2}r^2,
\end{equation}
where the constant $c$ is given by \eqref{c2}. The density force on the particle is $f_{eff}(r) = -u'_{eff}(r) $:
\begin{equation}
    f_{eff} = \Bigg(c + \frac{Ac^2}{2} \Bigg) - c^2r.
\end{equation}
At $r=0$
\begin{equation}
    f_{eff}(0) = c + \frac{Ac^2}{2}.
\end{equation}
The sign of the $f_{eff}(0)$ determines the behavior of the particle near the origin:
\begin{itemize}
    \item If $f_{eff}(0)>0 \rightarrow A>-\frac{2}{c}$, the force points outward, and the particle is repelled from the origin;
    \item If $f_{eff}(0)<0 \rightarrow A<-\frac{2}{c}$, the force points inward, and the particle is attracted toward the origin;
    \item If $f_{eff}(0)=0 \rightarrow A=-\frac{2}{c}$, the linear term vanishes, and the potential becomes harmonic.
\end{itemize}
We should stress that, when the force at the origin vanishes, the effective potential near $r=0$ becomes
\begin{eqnarray}
    u_{eff}(r) \approx \frac{c^2}{2}r^2.  
\end{eqnarray}
This is the potential of the harmonic oscillator. The origin is a point of stable equilibrium: if the particle is slightly displaced, a restoring force $f_{eff}(r) = -c^2 r$ pulls it back, causing it to oscillate around $r=0$. Locally, the particle remains attained near the center. For large distances, the potential does not diverge. Instead, it approaches a finite constant
\begin{equation}
u_\infty = \lim_{r\rightarrow\infty} u_{eff}(r) = -1-Ac.  \label{uinfty}
\end{equation}
We now evaluate the critical points in this case. The derivative of the effective potential \eqref{effn2} is:
\begin{equation}
    u'_{eff} = -ce^{-\frac{cr}{2}}\Bigg(e^{-\frac{cr}{2}}+\frac{Ac}{2}\Bigg).
\end{equation}
Critical points occur when $u'_{eff} = 0$. Since $e^{-\frac{cr}{2}}$ for finite $r=\tilde{r}$, we require
\begin{equation}
    e^{-\frac{c\tilde{r}}{2}}+\frac{Ac}{2} = 0 \implies e^{-\frac{c\tilde{r}}{2}} = -\frac{Ac}{2},\label{minu}
\end{equation}
then, the effective potential has extrema if $A < 0$. For $A \geq 0$, there are no extremal points for this potential. Then, it has extremal points when
\begin{equation}
    \tilde{r} = -\frac2c \ln\Bigg(-\frac{Ac}{2}\Bigg),\, \text{for}\, A<0. \label{rtilde}
\end{equation}
where $\tilde{r}$ is obtained by \eqref{minu}.
To evaluate whether points are minimum, we need to check the second derivative of the effective potential
\begin{equation}
    u''_{eff} = c^2e^{-\frac12 cr}\Bigg(e^{-\frac12 cr} + \frac14 Ac \Bigg).
\end{equation}
Then, for $A=-\frac2c$, the origin $r=0$ is a minimum. The result in \eqref{rtilde} gives also a minimum point, since $-\frac2c < A < 0$, as
\begin{equation}
    r_{min} = -\frac2c \ln\Bigg(-\frac{Ac}{2}\Bigg)> 0,\, \text{for}\, -\frac2c<A<0. 
\end{equation}
This regimen is characterized by a change in the sign of the force. Near the origin, the force is repulsive for $A>-\frac2c$. For large distances, the force becomes attractive when $A < 0$. The transition from repulsion to attraction gives rise to a local minimum at $r_{min} > 0$, corresponding to a stable equilibrium point. Then the conditions for equilibrium and stability of the spinning particle can be resumed as follows:
\begin{itemize}
    \item For $A < -\frac2c$, the origin is a region of stable equilibrium: small displacements result in a restoring force;
    \item For $A = -\frac2c$, the origin is a region of stable equilibrium (harmonic oscillator behavior);
    \item For $-\frac2c < A < 0$, there is no equilibrium at the origin, but a  region of stable equilibrium exists at $r_{min}>0$;
    \item For $A \geq 0$, no local minimum exists. The potential is strictly decreasing and bounded from below. There is no bounded motion; only scattering trajectories are possible. A particle coming from infinity is reflected and returns to infinity, never remaining in a finite region.
\end{itemize}
Because the potential approaches a finite constant at infinity, a particle can escape to infinity only if its total energy $E > -1 - Ac$. For energies below this threshold, the particle is confined to a finite region. However, the existence of the confinement depends on the presence of a local minimum. Thus, confinement occurs only for $A<0$. In all these cases, the particle oscillates around a stable equilibrium point: either at the origin ($A < -\frac2c$) or at a finite distance ($-\frac2c < A < 0$). For $A \geq 0$, the system supports only unbounded trajectories and confinement is not possible.

\begin{figure}[!ht]
    \centering
    \includegraphics[width=0.95\textwidth]{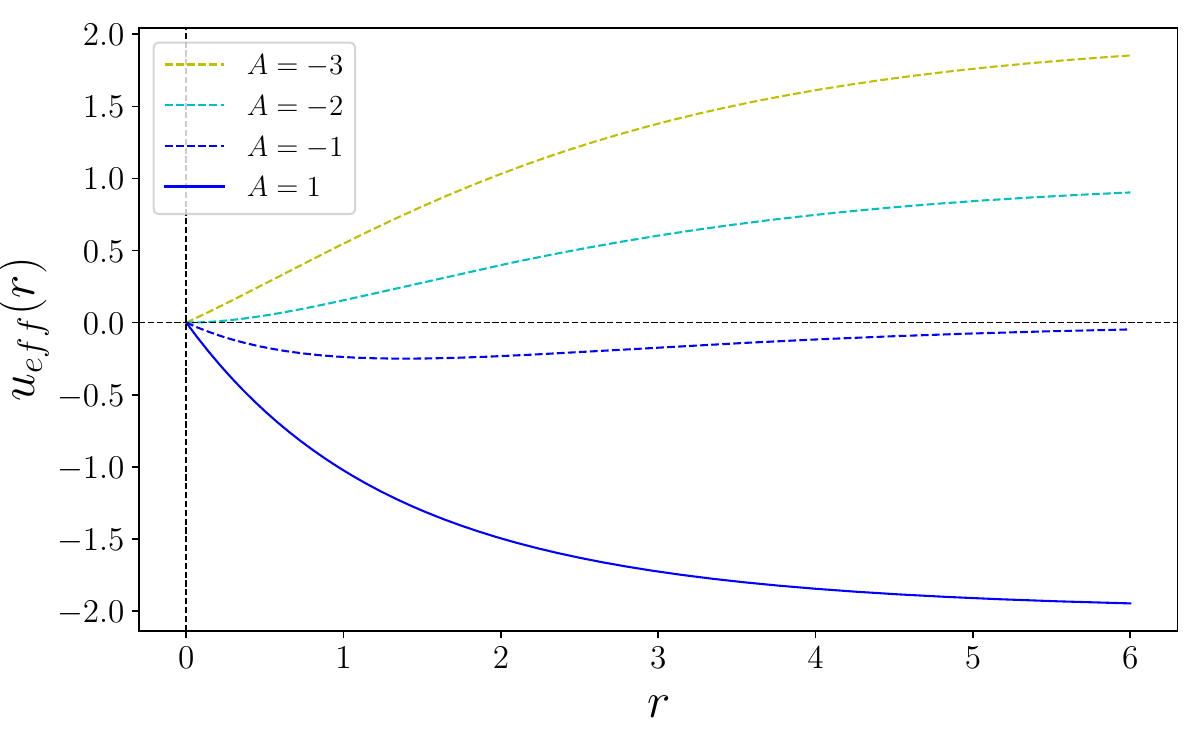}
        \caption{The effective Hedgehog potential for the spin 1/2 particles. We use $c=1$.}
        \label{fig:PotHedgehogspin1/2NOVO}
\end{figure}
The Figure~\ref{fig:PotHedgehogspin1/2NOVO} reproduce the behavior of equation (\ref{effn2}), setting $c=1$  and different values of $A [-3,-2,-1,1]$.

\newpage

\subsection{ Spinning Particle in Codimension \texorpdfstring{$n\geq3$}{n>=3}: the bulk scalar field as topological defect}

Now we will adopt the metric \eqref{metriccdn} to describe the background in codimension $n\geq3$, where a global topological defect is generated by a bulk scalar field. Then, the Christoffel symbols are given by \eqref{gammast}, \eqref{gammanw} and \eqref{gammaij}. The Riemann Tensor has the components given below:
\begin{eqnarray}
    &&R^{\mu}_{\nu\kappa\tau}=\frac{1}{4}\frac{\sigma'^{2}( r )}{\sigma^2(r)}\Big[\delta^{\mu}_{\tau}g_{\nu\kappa}-\delta^{\mu}_{\kappa}g_{\nu\tau}\Big];\label{ricgen1}\\
    &&R^{\mu}_{ r  r \tau}=\left[\frac{1}{2}\frac{\sigma''( r )}{\sigma( r )}-\frac{1}{4}\frac{\sigma'^{2}( r )}{\sigma^{2}( r )}\right]\delta^{\mu}_{\tau};\\
    &&R^{\mu}_{ij\nu}=\frac{1}{4}\frac{\sigma'^2( r )}{\sigma^2( r )}\gamma(r)\delta^{\mu}_{\nu}\xi_{ij};\\
    &&R^{ r }_{\mu\nu  r }=\left[\frac{1}{2}\sigma''( r )-\frac{1}{4}\frac{\sigma'^{2}( r )}{\sigma( r )}\right]\eta_{\mu\nu};\\
    &&R^{ r }_{ij  r }=\left[\frac{1}{2}\gamma''( r )-\frac{1}{4}\frac{\gamma'^{2}( r )}{\gamma( r )}\right]\xi_{ij};\label{ricgen5}\\
    &&R^{i}_{jkl}=\tilde{R}^{i}_{jkl};\\
    &&R^{i}_{ r  r  l}=\left[\frac{1}{2}\frac{\gamma''( r )}{\gamma( r )}-\frac{1}{4}\frac{\gamma'^{2}( r )}{\gamma^{2}( r )}\right]\delta^{i}_{l}.
\end{eqnarray}
The solutions \eqref{constp5} and \eqref{psipu} obtained in the previous section are valid here. Then, following the same steps, using \eqref{mot1gen}, \eqref{mot2gen}, \eqref{constp5}, \eqref{psipu}, \eqref{ricgen1}-\eqref{ricgen5} in \eqref{506}, we get the conserved quantities in the directions $x^\mu$ and $r$ given by:
\begin{eqnarray}
 &&\frac{d}{d\tau}(\sigma \dot{x}^\mu + \frac i2 \psi^r_k \psi^\mu_k\sigma') = 0;\label{consxgen} \\
 && \frac{d}{d\tau}[\sigma(\dot{r}^2 + m^2)-im\psi^5_k \psi^r_k(0)\sigma'\sigma^{-\frac12}] = 0. \label{consrgen}
\end{eqnarray}

For codimensions greater than or equal to three, the conserved momentum and the effective potential reduce to the same expression obtained in the codimension-$2$ case:
\begin{eqnarray}
    &&  p^\mu = \sigma \dot{x}^\mu + \frac i2 \psi^r_k \psi^\mu_k\sigma' ,\\
    && u_{eff} =  \sigma - A\Big[\sigma'\sigma^{-\frac12}-\sigma'(0)\Big] -1,\label{ueffs2n3}
\end{eqnarray}
where $\sigma(r)$ is given by \eqref{sigman3+} with a constant $c$ equal to \eqref{cn3}.  This expression is identical to the one analyzed in section~\ref{spincd2}. Therefore, the entire discussion regarding the behavior near the origin, the existence of local minima, and the conditions for classical confinement applies without modification.

In particular, the effective force in the origin is given by $f_{eff}(0) = c + \frac{Ac^2}{2}$. The following regimen is obtained:
\begin{itemize}
    \item If $A < -\frac2c$, the particle is attracted to the origin, which is a region of stable equilibrium;
    \item If $A = -\frac2c$, the potential reduces to that of a harmonic oscillator at the origin;
    \item If $-\frac2c < A < 0$, there is a local minimum at $r_{min} >0$, where the particle oscillates for low energies;
    \item For $A \geq 0$, no local minimum exists. The potential is strictly decreasing and supports only scattering trajectories.
\end{itemize}
Thus, confinement occurs only for 
$A < 0$, exactly as in the codimension $2$ case. The parameter $A$ controls the transition between attraction, harmonic confinement, and repulsion, with the same physical consequences discussed in section~\ref{spincd2}.

\begin{figure}[htbp]
    \centering
    \begin{subfigure}[b]{0.45\textwidth}
        \centering
        \includegraphics[width=1.0\textwidth]{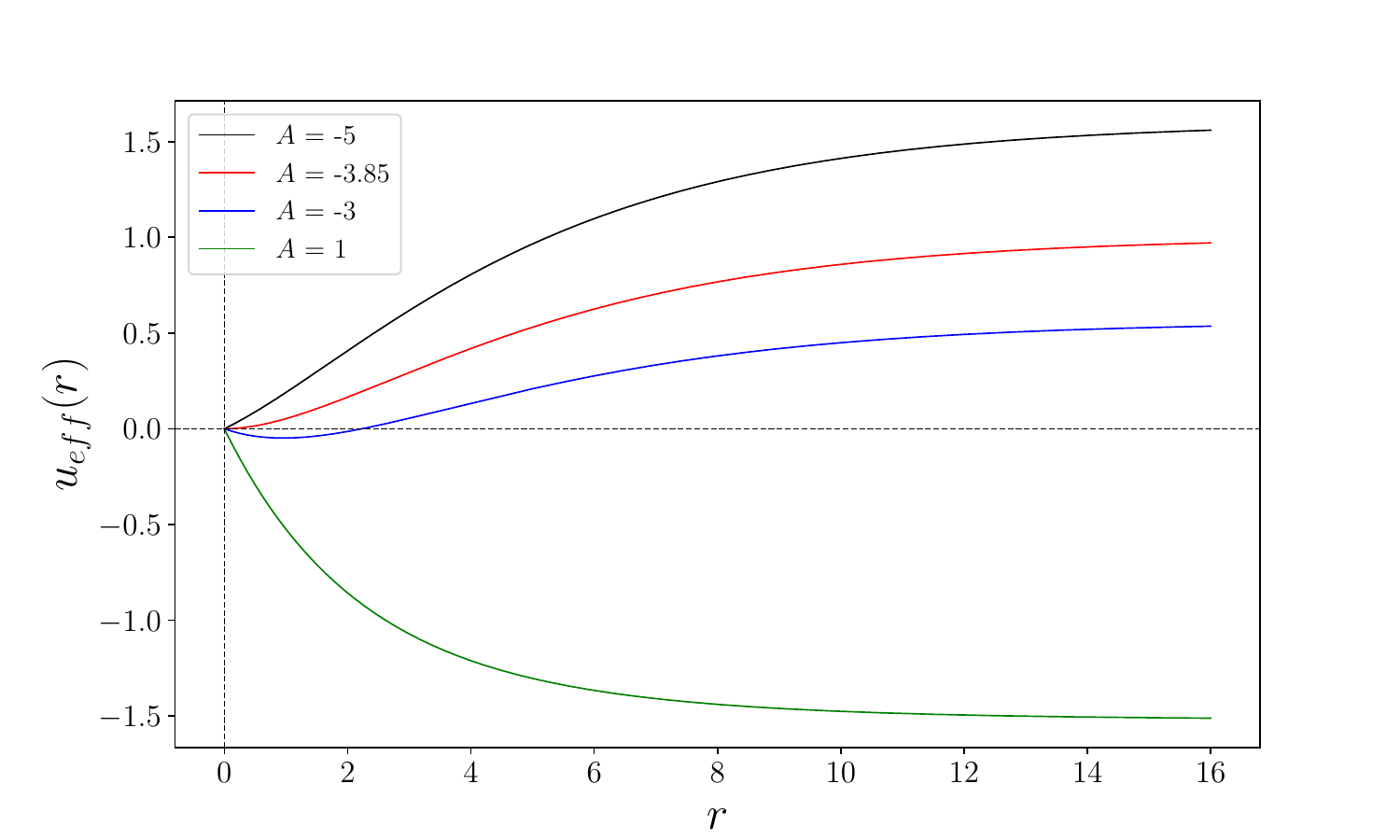}
        \caption{The effective Hedgehog potential for the spinning particles $N=1,2$ in codimension $3$ to $ c_{n=3} = 0.52. $}
        \label{subc3}
    \end{subfigure}
    \hfill 
    \begin{subfigure}[b]{0.45\textwidth}
        \centering
        \includegraphics[width=1.0\textwidth]{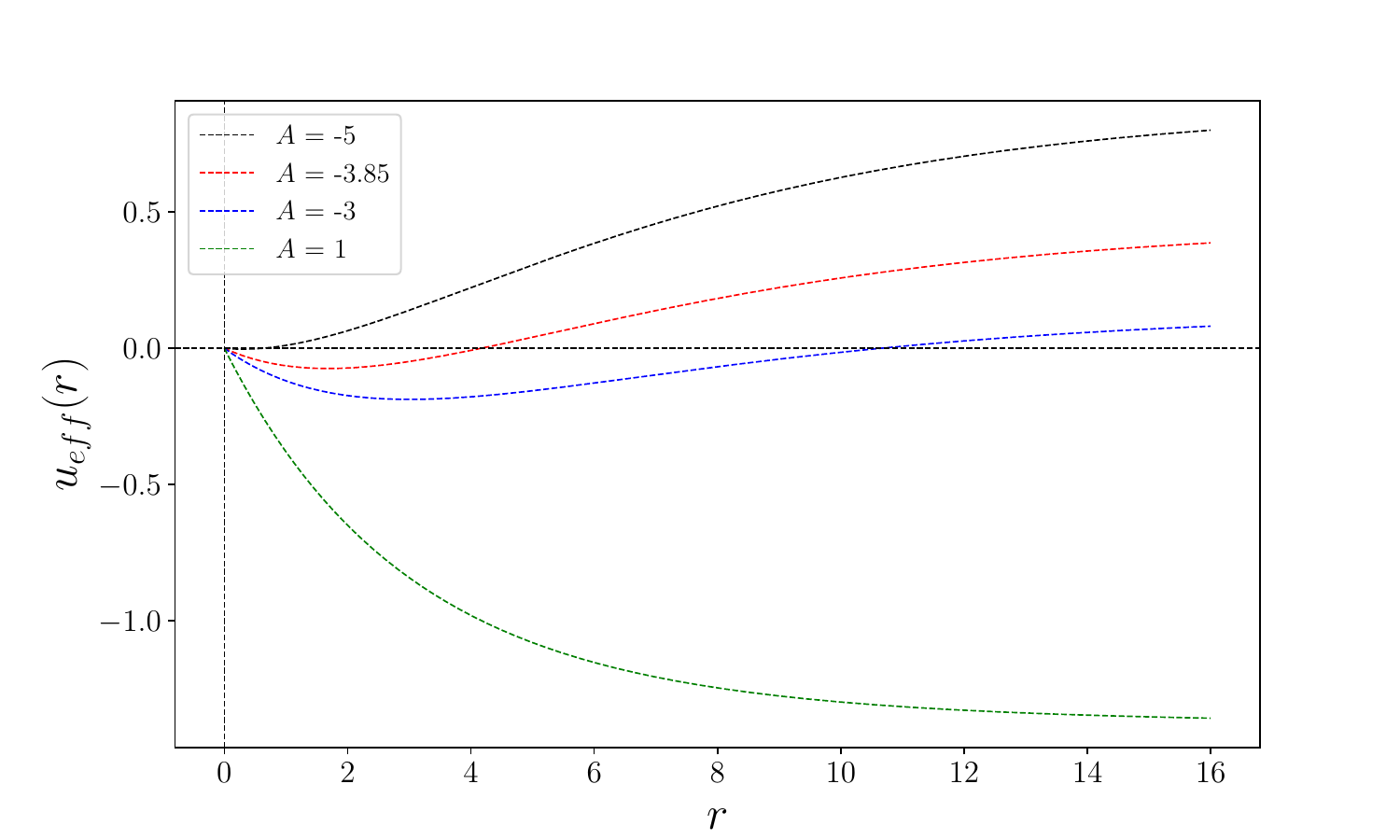}
        \caption{The effective Hedgehog potential for the spinning particles $N=1,2$ in codimension $5$, to $ c_{n=5} = 0.38. $}
        \label{subc5}
    \end{subfigure}
    
    \caption{Plot of effective Hedgehog potential for spinning particles $N=1,2$ in codimensions $3$\eqref{subc3} and $5$\eqref{subc5} to the same values of the parameter $A[-5, -3.85, -3, 1]$. }
    \label{fig:lado_a_lado}
\end{figure}

Two points must be considered here, as discussed in the previous section regarding the choice of background. The first one is that if we take exactly the same background, that is, if we choose the same $c$, we will obtain the same behavior for both cases, as seen in Figure~\ref{fig:PotHedgehogspin1/2NOVO}. On the other hand (the second point), by adjusting the parameter $c$  according to the choice of $n$, the $c_{n = 3} = 0.52$ and  $c_{n=5} = 0.38$, fixing $-\Lambda_{D} /M_{D}^{n+2} = 1$ and keeping the choice of parameter $A$ as shown in the Figure~\ref{fig:lado_a_lado}, again we see a more pronounced behavior.

\newpage
\section{Conclusions and Perspectives} \label{sec:conclusion_perspective}

In this manuscript, we study the behavior of the spinless free particle, the $N=1$ and $N=2$ spinning particle in a braneworld scenario with codimension $n$. The main question we want to answer is related to the impact of codimension on confinement of particles to the membrane. We graphically analyze the effective potentials for  codimension $n=2$, with a local string-like topological defect \eqref{metriccd2} and for codimension $n\geq3$ where a global topological defect is generated by a bulk
scalar field \eqref{metricgheta}. The effective potentials for codimension $2$ and codimension $n \geq 3$ both for the spinless particle and the spinning particle $N=1,2$ are given by the pairs [\eqref{ueffcd2dens}, \eqref{ueffcd3+}] and [\eqref{effn2}, \eqref{ueffs2n3}], respectively, where $\sigma(r) = e^{-cr}$ and the constant $c$ given by \eqref{c2} and \eqref{cn3}, respectively.
The behavior of the spinless particle can be determined by analyzing the effective potential
\[
u_{eff}(r) = e^{-cr} - 1.
\]
This analysis was carried out for the effective potential that arises from the codimension $2$ model, which is the same structure that appears in higher codimensions $n \geq 3$. Therefore, the entire discussion regarding equilibrium, stability, and confinement applies directly to those cases. Near the origin, the force is repulsive, with $f_{eff}(0) = c > 0$. The potential decreases monotonically from $u_{eff}(0) = 0$ to $u_{eff}(r \rightarrow \infty) = -1$, with no critical points for $r>0$. Consequently, there is no stable equilibrium anywhere in the system. Because the potential approaches a finite constant at infinity, a particle can escape to infinity only if its total energy $E \geq -1$. For energies below this threshold, there is no classical motion. For energies above this value, the particle follows scattering trajectories, coming from infinity and returning to infinity after reflection. Thus, confinement is not possible.

For $N=1$ spinning particles, which represents spin $1/2$ particles, and $N=2$ spinning particle, which represents spin $ 1$ particles, the behavior is completely determined by the value of $A$. It controls the force at the origin, the existence of local minima, and the nature of the particle motion. Again, we analyze the effective potential that arises from the problem of codimension $2$. The same structure appears in higher codimension $n\geq3$. Therefore, the entire discussion regarding equilibrium, stability, and confinement applies directly to those cases without modification. We find by evaluating the effective potential
\[
u_{eff} =  e^{-cr} - Ac\Big[e^{-\frac12 cr}-1\Big] -1,
\]
near the origin, that the particle is subject to the force $f_{eff} = c + \frac{Ac}{2}.$
For $A<-\frac2c$, the particle is attracted to the origin, which acts as a region of stable equilibrium. For $A=-\frac2c$, the potential reduces to a harmonic oscillator with stable equilibrium at the origin. For $A> -\frac2c$ the particle is repelled from the origin. In the intermediate regime $-\frac2c < A < 0$, a local minimum appears at $r_{min}$, creating a potential well away from the origin. For $A\geq0$, there is no local minimum: the potential is strictly decreasing and supports only unbounded trajectories.

Because the potential approaches a finite constant $u_\infty = -1-Ac$ as $r\rightarrow\infty$, a particle can escape to infinity only if its energy exceeds this value. However, confinement occurs only when there is a local minimum, i.e., for $A<0$. In these cases, the particle oscillates around a stable equilibrium point — either at the origin ($A\leq-\frac2c$) or at a finite distance ($-\frac2c < A < 0$). For $A\geq0$, confinement is not possible: the system exhibits only scattering states. Thus, the parameter $A$ determines the transition between attraction, harmonic confinement, repulsion with a distant well, and purely repulsive scattering recovering the spinless case when $A = 0$. The existence of a local minimum in the effective potential is the condition for confinement in this classical system, for codimension $n \geq 2$.

As shown in codimension $1$ \cite{Dahia:2007ep, Souza:2019jqz, deSouza:2025wzv}, the free particles and the spinning particles are not confined to the brane, but the spinning particles are confined in a region near the brane. Moreover, for codimension $n \geq 2$, the spinless particles do not remain attached to the membrane, while for the spinning particles the confinement is possible on the membrane, depending on the value of the parameter $A$. Then, the spinning particle can be attached to the brane or yet present us the satellite behavior obtained in \cite{Souza:2019jqz, deSouza:2025wzv}.

For codimension $n \geq 2$ in these backgrounds, the confinement of free particles remains an open problem. The spin $1/2$ particles ($N=1$) and $p$-forms ($N=2$) can be trapped on the membrane or localized near it as satellite objects \cite{Souza:2019jqz, deSouza:2025wzv}, depending on the value of parameter $A$ and the total energy along the radial extra dimension $E < u_\infty$. This result directly affects the scalar field, spinor and $p$-form interpretation and localization of fields on the membrane just as verified in the usual models.  

The study of the $N=4$ superparticle may be interesting because, after quantization, a gravity interpretation can be built \cite{Bonezzi:2018box}. In this case, the question of gravity confinement can be addressed in this context. Another interesting direction is related to particles in the gravitational backgrounds of intersecting membranes \cite{Gauy:2022yxj,Gauy:2022xsc}, such ones that could mimic the results given here due to different codimension. These are nice questions to future research.

\acknowledgments

I. M. Macêdo and F. E. A. de Souza are thankful for the financial support provided by the Fundação Cearense de Apoio ao Desenvolvimento Científico e Tecnológico (FUNCAP) through processes n{$^\circ$} 31052.000190/2025-40 and FPD-$0213$-$00349.01.01/23$, respectively.


\bibliographystyle{JHEP}
\bibliography{biblio.bib}

@book{Polchinski:1998rq,
    author = "Polchinski, J.",
    title = "{String theory. Vol. 1: An introduction to the bosonic string}",
    doi = "10.1017/CBO9780511816079",
    isbn = "978-0-511-25227-3, 978-0-521-67227-6, 978-0-521-63303-1",
    publisher = "Cambridge University Press",
    series = "Cambridge Monographs on Mathematical Physics",
    month = "12",
    year = "2007"
}

@book{Polchinski:1998rr,
    author = "Polchinski, J.",
    title = "{String theory. Vol. 2: Superstring theory and beyond}",
    doi = "10.1017/CBO9780511618123",
    isbn = "978-0-511-25228-0, 978-0-521-63304-8, 978-0-521-67228-3",
    publisher = "Cambridge University Press",
    series = "Cambridge Monographs on Mfathematical Physics",
    month = "12",
    year = "2007"
}

@inbook{Liu:2017gcn,
    author = "Liu, Yu-Xiao",
    title = "{Introduction to Extra Dimensions and Thick Braneworlds}",
    eprint = "1707.08541",
    archivePrefix = "arXiv",
    primaryClass = "hep-th",
    doi = "10.1142/9789813237278_0008",
    year = "2018",
    booktitle = "Memorial Volume for Yi-Shi Duan"
}

@article{Olasagasti:2000gx,
    author = "Olasagasti, Itsaso and Vilenkin, Alexander",
    title = "{Gravity of higher dimensional global defects}",
    eprint = "hep-th/0003300",
    archivePrefix = "arXiv",
    doi = "10.1103/PhysRevD.62.044014",
    journal = "Phys. Rev. D",
    volume = "62",
    pages = "044014",
    year = "2000"
}

@article{Gherghetta:2000qi,
    author = "Gherghetta, Tony and Shaposhnikov, Mikhail E.",
    title = "{Localizing gravity on a string - like defect in six-dimensions}",
    eprint = "hep-th/0004014",
    archivePrefix = "arXiv",
    reportNumber = "UNIL-IPT-00-08",
    doi = "10.1103/PhysRevLett.85.240",
    journal = "Phys. Rev. Lett.",
    volume = "85",
    pages = "240--243",
    year = "2000"
}

@article{Berkovits:2002zk,
    author = "Berkovits, Nathan",
    editor = "Bachas, C. and Gava, E. and Maldacena, Juan Martin and Narain, K. S. and Randjbar-Daemi, S.",
    title = "{ICTP lectures on covariant quantization of the superstring}",
    eprint = "hep-th/0209059",
    archivePrefix = "arXiv",
    reportNumber = "IFT-P-063-2002",
    journal = "ICTP Lect. Notes Ser.",
    volume = "13",
    pages = "57--107",
    year = "2003"
}

@article{Berkovits:2001rb,
    author = "Berkovits, Nathan",
    title = "{Covariant quantization of the superparticle using pure spinors}",
    eprint = "hep-th/0105050",
    archivePrefix = "arXiv",
    reportNumber = "IFT-P-000-2001",
    doi = "10.1088/1126-6708/2001/09/016",
    journal = "JHEP",
    volume = "09",
    pages = "016",
    year = "2001"
}

@article{Randall:1999vf,
    author = "Randall, Lisa and Sundrum, Raman",
    title = "{An Alternative to compactification}",
    eprint = "hep-th/9906064",
    archivePrefix = "arXiv",
    reportNumber = "MIT-CTP-2874, PUPT-1867, BUHEP-99-13",
    doi = "10.1103/PhysRevLett.83.4690",
    journal = "Phys. Rev. Lett.",
    volume = "83",
    pages = "4690--4693",
    year = "1999"
}

@article{Randall:1999ee,
    author = "Randall, Lisa and Sundrum, Raman",
    title = "{A Large mass hierarchy from a small extra dimension}",
    eprint = "hep-ph/9905221",
    archivePrefix = "arXiv",
    reportNumber = "MIT-CTP-2860, PUPT-1860, BUHEP-99-9",
    doi = "10.1103/PhysRevLett.83.3370",
    journal = "Phys. Rev. Lett.",
    volume = "83",
    pages = "3370--3373",
    year = "1999"
}

@article{Mukhopadhyaya:2007jn,
    author = "Mukhopadhyaya, Biswarup and Sen, Somasri and SenGupta, Soumitra",
    title = "{Bulk antisymmetric tensor fields in a Randall-Sundrum model}",
    eprint = "0709.3428",
    archivePrefix = "arXiv",
    primaryClass = "hep-th",
    doi = "10.1103/PhysRevD.76.121501",
    journal = "Phys. Rev. D",
    volume = "76",
    pages = "121501",
    year = "2007"
}

@article{Germani:2004jf,
    author = "Germani, Cristiano and Kehagias, Alex",
    title = "{Higher-spin fields in braneworlds}",
    eprint = "hep-th/0411269",
    archivePrefix = "arXiv",
    reportNumber = "DAMTP-2004-140",
    doi = "10.1016/j.nuclphysb.2005.07.027",
    journal = "Nucl. Phys. B",
    volume = "725",
    pages = "15--44",
    year = "2005"
}

@article{Alencar:2010mi,
    author = "Alencar, G. and Tahim, M. O. and Landim, R. R. and Muniz, C. R. and Costa Filho, R. N.",
    title = "{Bulk Antisymmetric tensor fields coupled to a dilaton in a Randall-Sundrum model}",
    eprint = "1005.1691",
    archivePrefix = "arXiv",
    primaryClass = "hep-th",
    doi = "10.1103/PhysRevD.82.104053",
    journal = "Phys. Rev. D",
    volume = "82",
    pages = "104053",
    year = "2010"
}

@article{Tahim:2008ka,
    author = "Tahim, M. O. and Cruz, W. T. and Almeida, C. A. S.",
    title = "{Tensor gauge field localization in branes}",
    eprint = "0808.2199",
    archivePrefix = "arXiv",
    primaryClass = "hep-th",
    doi = "10.1103/PhysRevD.79.085022",
    journal = "Phys. Rev. D",
    volume = "79",
    pages = "085022",
    year = "2009"
}

@article{Fu:2016vaj,
    author = "Fu, Chun-E and Zhong, Yuan and Xie, Qun-Ying and Liu, Yu-Xiao",
    title = "{Localization and mass spectrum of $q-$form fields on branes}",
    eprint = "1601.07118",
    archivePrefix = "arXiv",
    primaryClass = "hep-th",
    doi = "10.1016/j.physletb.2016.03.069",
    journal = "Phys. Lett. B",
    volume = "757",
    pages = "180--186",
    year = "2016"
}

@article{Lu:2024gmx,
    author = "Lu, Yong-Tao and Guo, Heng and Wei, Qun and Fu, Chun-E",
    title = "{Localization mechanism of q-form field on the braneworld by coupling with gravity}",
    eprint = "2401.11688",
    archivePrefix = "arXiv",
    primaryClass = "hep-th",
    doi = "10.1103/PhysRevD.111.085025",
    journal = "Phys. Rev. D",
    volume = "111",
    number = "8",
    pages = "085025",
    year = "2025"
}

@article{Guo:2023mki,
    author = "Guo, Heng and Lu, Yong-Tao and Wang, Cai-Ling and Sun, Yue",
    title = "{Localization of scalar field on the brane-world by coupling with gravity}",
    eprint = "2310.01451",
    archivePrefix = "arXiv",
    primaryClass = "hep-th",
    doi = "10.1007/JHEP06(2024)114",
    journal = "JHEP",
    volume = "06",
    pages = "114",
    year = "2024"
}

@article{Wan:2023usr,
    author = "Wan, Jun-Jie and Liu, Yu-Xiao",
    title = "{Localization of spinor fields in higher-dimensional braneworlds}",
    eprint = "2303.06278",
    archivePrefix = "arXiv",
    primaryClass = "hep-th",
    doi = "10.1007/JHEP12(2023)033",
    journal = "JHEP",
    volume = "12",
    pages = "033",
    year = "2023"
}

@article{Dahia:2007ep,
    author = "Dahia, F. and Romero, C.",
    title = "{Confinement and stability of the motion of test particles in thick branes}",
    eprint = "gr-qc/0702011",
    archivePrefix = "arXiv",
    doi = "10.1016/j.physletb.2007.06.004",
    journal = "Phys. Lett. B",
    volume = "651",
    pages = "232--238",
    year = "2007"
}

@article{Oda:2000zc,
    author = "Oda, Ichiro",
    title = "{Localization of matters on a string - like defect}",
    eprint = "hep-th/0006203",
    archivePrefix = "arXiv",
    reportNumber = "EDO-EP-30",
    doi = "10.1016/S0370-2693(00)01284-3",
    journal = "Phys. Lett. B",
    volume = "496",
    pages = "113--121",
    year = "2000"
}

@article{Gherghetta:2000jf,
    author = "Gherghetta, Tony and Roessl, Ewald and Shaposhnikov, Mikhail E.",
    title = "{Living inside a hedgehog: Higher dimensional solutions that localize gravity}",
    eprint = "hep-th/0006251",
    archivePrefix = "arXiv",
    reportNumber = "UNIL-IPT-00-14",
    doi = "10.1016/S0370-2693(00)00979-5",
    journal = "Phys. Lett. B",
    volume = "491",
    pages = "353--361",
    year = "2000"
}

@article{Howe:1989vn,
    author = "Howe, Paul S. and Penati, Silvia and Pernici, Mario and Townsend, Paul K.",
    title = "{A Particle Mechanics Description of Antisymmetric Tensor Fields}",
    reportNumber = "CERN-TH-5305-89",
    doi = "10.1088/0264-9381/6/8/012",
    journal = "Class. Quant. Grav.",
    volume = "6",
    pages = "1125",
    year = "1989"
}

@article{Souza:2019jqz,
    author = "Souza, F. E. A. and Freitas, L. F. F. and Alencar, G. and Landim, R. R.",
    title = "{Confinement of bosonic and spinning particles in braneworlds}",
    eprint = "1906.11665",
    archivePrefix = "arXiv",
    primaryClass = "hep-th",
    doi = "10.1209/0295-5075/133/50001",
    journal = "EPL",
    volume = "133",
    number = "5",
    pages = "50001",
    year = "2021"
}

@article{Berezin:1976eg,
    author = "Berezin, F. A. and Marinov, M. S.",
    title = "{Particle Spin Dynamics as the Grassmann Variant of Classical Mechanics}",
    reportNumber = "ITEP-43-1976",
    doi = "10.1016/0003-4916(77)90335-9",
    journal = "Annals Phys.",
    volume = "104",
    pages = "336",
    year = "1977"
}

@article{Casalbuoni:1975hx,
    author = "Casalbuoni, R.",
    title = "{Relativity and Supersymmetries}",
    reportNumber = "Print-76-0053 (FLORENCE)",
    doi = "10.1016/0370-2693(76)90044-7",
    journal = "Phys. Lett. B",
    volume = "62",
    pages = "49--50",
    year = "1976"
}

@article{Casalbuoni:1975bj,
    author = "Casalbuoni, R.",
    title = "{On the Quantization of Systems with Anticommutating Variables}",
    reportNumber = "Print-75-1061 (FLORENCE)",
    doi = "10.1007/BF02748689",
    journal = "Nuovo Cim. A",
    volume = "33",
    pages = "115",
    year = "1976"
}

@article{Brink:1976uf,
    author = "Brink, L. and Di Vecchia, P. and Howe, Paul S.",
    title = "{A Lagrangian Formulation of the Classical and Quantum Dynamics of Spinning Particles}",
    reportNumber = "NBI-HE-76-8",
    doi = "10.1016/0550-3213(77)90364-9",
    journal = "Nucl. Phys. B",
    volume = "118",
    pages = "76--94",
    year = "1977"
}

@article{deSouza:2025wzv,
    author = "de Souza, F. E. A. and Tahim, M. O. and de Oliveira Junior, R. I. and Macedo, I. M.",
    title = "{A satellite $N=2$ superparticle in extra dimensions}",
    eprint = "2508.13970",
    archivePrefix = "arXiv",
    primaryClass = "hep-th",
    doi = "10.1140/epjc/s10052-025-15195-6",
    journal = "Eur. Phys. J. C",
    volume = "85",
    number = "12",
    pages = "1446",
    year = "2025"
}

@article{Howe:1988ft,
    author = "Howe, Paul S. and Penati, Silvia and Pernici, Mario and Townsend, Paul K.",
    title = "{Wave Equations for Arbitrary Spin From Quantization of the Extended Supersymmetric Spinning Particle}",
    reportNumber = "IFUM-345/FT",
    doi = "10.1016/0370-2693(88)91358-5",
    journal = "Phys. Lett. B",
    volume = "215",
    pages = "555--558",
    year = "1988"
}

@article{Rivelles:1990dq,
    author = "Rivelles, Victor O. and Sandoval, Jr., L.",
    title = "{BRST quantization of relativistic spinning particles with a Chern-Simons term}",
    reportNumber = "IFUSP-P-874",
    doi = "10.1088/0264-9381/8/8/022",
    journal = "Class. Quant. Grav.",
    volume = "8",
    pages = "1605--1611",
    year = "1991"
}

@article{Brink:1976sz,
    author = "Brink, L. and Deser, Stanley and Zumino, B. and Di Vecchia, P. and Howe, Paul S.",
    editor = "Salam, A. and Sezgin, E.",
    title = "{Local Supersymmetry for Spinning Particles}",
    reportNumber = "CERN-TH-2208",
    doi = "10.1016/0370-2693(76)90115-5",
    journal = "Phys. Lett. B",
    volume = "64",
    pages = "435",
    year = "1976",
    note = "[Erratum: Phys.Lett.B 68, 488 (1977)]"
}

@article{Gershun:1979fb,
    author = "Gershun, V. D. and Tkach, V. I.",
    title = "{CLASSICAL AND QUANTUM DYNAMICS OF PARTICLES WITH ARBITRARY SPIN}",
    journal = "JETP Lett.",
    volume = "29",
    pages = "288--291",
    year = "1979"
}

@article{Rietdijk:1989qa,
    author = "Rietdijk, R. H. and van Holten, J. W.",
    title = "{Generalized Killing Equations and Symmetries of Spinning Space}",
    reportNumber = "NIKHEF-H/89-8",
    doi = "10.1088/0264-9381/7/2/017",
    journal = "Class. Quant. Grav.",
    volume = "7",
    pages = "247",
    year = "1990"
}

@article{Brink:1976sc,
    author = "Brink, L. and Di Vecchia, P. and Howe, Paul S.",
    title = "{A Locally Supersymmetric and Reparametrization Invariant Action for the Spinning String}",
    reportNumber = "GOTEBORG-76-22",
    doi = "10.1016/0370-2693(76)90445-7",
    journal = "Phys. Lett. B",
    volume = "65",
    pages = "471--474",
    year = "1976"
}

@article{Siegel:1988ru,
    author = "Siegel, W.",
    title = "{Conformal Invariance of Extended Spinning Particle Mechanics}",
    reportNumber = "ITP-SB-88-41",
    doi = "10.1142/S0217751X88001132",
    journal = "Int. J. Mod. Phys. A",
    volume = "3",
    pages = "2713--2718",
    year = "1988"
}

@article{Bonezzi:2018box,
    author = "Bonezzi, R. and Meyer, A. and Sachs, I.",
    title = "{Einstein gravity from the $ \mathcal{N}=4 $ spinning particle}",
    eprint = "1807.07989",
    archivePrefix = "arXiv",
    primaryClass = "hep-th",
    doi = "10.1007/JHEP10(2018)025",
    journal = "JHEP",
    volume = "10",
    pages = "025",
    year = "2018"
}

@article{Gauy:2022yxj,
    author = "Gauy, Henrique Matheus and Bernardini, Alex E.",
    title = "{Gravity localization on intersecting thick braneworlds}",
    eprint = "2209.09215",
    archivePrefix = "arXiv",
    primaryClass = "hep-th",
    doi = "10.1103/PhysRevD.106.084003",
    journal = "Phys. Rev. D",
    volume = "106",
    number = "8",
    pages = "084003",
    year = "2022"
}

@article{Gauy:2022xsc,
    author = "Gauy, Henrique Matheus and Bernardini, Alex E.",
    title = "{(5+1)-dimensional analytical braneworld models: Intersecting thick branes}",
    eprint = "2201.01284",
    archivePrefix = "arXiv",
    primaryClass = "hep-th",
    doi = "10.1103/PhysRevD.105.024068",
    journal = "Phys. Rev. D",
    volume = "105",
    number = "2",
    pages = "024068",
    year = "2022"
}


\end{document}